\documentclass[aps,floats,twocolumn,prl,nofootinbib]{revtex4-2}
\usepackage{amsfonts,amsmath,amssymb,ascmac,bm,tensor}
\usepackage{fnpct} 
\usepackage{comment}
\usepackage{ifpdf}
\usepackage{slashed}
\usepackage{color}
\usepackage[mathscr]{eucal}
\usepackage[utf8]{inputenc}
\usepackage{physics}
\usepackage{cancel}
\usepackage{subcaption}
\usepackage{soul}
\usepackage{simpler-wick}
\usepackage[breaklinks=true]{hyperref}
\def\r{\right}

\begin{document}

\title{Ward Identity Constraints on Loop Corrections in Non-Attractor Inflation}

\author{Cheng-Jun Fang$^{1,2}$}
\email{fangchengjun@itp.ac.cn}
\author{Zhen-Hong Lyu$^{1,2}$}
\email{lyuzhenhong@itp.ac.cn}
\author{Chao Chen$^{3}$}
\email{cchao012@just.edu.cn}
\author{Zong-Kuan Guo$^{1,2,4}$}
\email{guozk@itp.ac.cn}

    \affiliation{$^1$Institute of Theoretical Physics, Chinese Academy of Sciences, P.O. Box 2735, Beijing 100190, China}
    \affiliation{$^2$School of Physical Sciences, University of Chinese Academy of Sciences, No.19A Yuquan Road, Beijing 100049, China}
    \affiliation{$^3$School of Science, Jiangsu University of Science and Technology, Zhenjiang 212100, China}
    \affiliation{$^4$School of Fundamental Physics and Mathematical Sciences, Hangzhou Institute for Advanced Study, University of Chinese Academy of Sciences, Hangzhou 310024, China}

\begin{abstract}
The conservation of super-horizon curvature perturbations in strongly interacting inflationary models, particularly in the presence of quantum-loop corrections, remains a topic of active debate. We found that this conservation is essentially a direct consequence of the symmetry in perturbation theory. We demonstrate that the associated Ward identity imposes strict non-perturbative constraints on the infrared power spectrum. This finding provides a rigorous, symmetry-based framework for understanding nonlinear quantum fluctuations in the primordial universe.
\end{abstract}

\maketitle

\textit{\textbf{Introduction.}} 
The conservation of superhorizon perturbations is a cornerstone of inflationary cosmology, bridging the physics of the early universe with late-time observables like the cosmic microwave background. At linear level, the constancy of the comoving curvature perturbation $\zeta$ has been rigorously established in seminal work~\cite{Bardeen1980,KodamaSasaki1984,Wands:2000dp,Weinberg:2003sw}. 
Furthermore, research demonstrates that this conservation persists under certain conditions, even when non-linearities are incorporated, as shown in classical analyses~\cite{Lyth:2004gb,Langlois:2005qp,Rigopoulos:2003ak,Abolhasani:2019cqw} and in quantum aspects~\cite{Assassi:2012et,Senatore:2012ya,Tanaka:2015aza}. 
However, this conservation confronts a fundamental limitation in the case of strong interactions. Existing proofs conventionally presuppose weak non-Gaussianity, effectively constraining the dynamics to regimes where linear approximations remain valid. Such frameworks systematically neglect scenarios most susceptible to conservation violations, particularly those involving significant loop corrections.

In this letter, we point out that this infrared conservation is essentially a conservation law guaranteed by symmetry, and we have extended its scope of application to quantum scenarios with strong nonlinear effects. In fact, whether this conservation law can be extended to such scenarios has been subject to much skepticism recently, sparked by models of ultra-slow-roll (USR) inflation for producing primordial black holes~\cite{Motohashi:2017kbs,Kinney:1997ne,Inoue:2001zt,Kinney:2005vj,Martin:2012pe,Liu:2020oqe,Ozsoy:2019lyy}. The seminal work~\cite{Kristiano:2022maq} argued that one-loop corrections to $\zeta$ in USR is time-evolving in the superhorizon limit, and could even dominate over tree-level contributions.
Subsequent work, spanning various methods, robustly affirms large, non-decaying super-Hubble loop corrections.~\cite{Kristiano:2023scm,Firouzjahi:2023aum,Firouzjahi:2023bkt,kristianoComparingSharpSmooth2024,Choudhury:2023vuj,Franciolini:2023agm,Iacconi:2023ggt,Davies:2023hhn,Maity:2023qzw,Tasinato:2023ukp,Cheng:2023ikq,Ballesteros:2024zdp,Sheikhahmadi:2024peu}. 
Such a violation of the super-Hubble conservation of curvature perturbations would erode a cornerstone of inflationary theory, undermining the standard cosmological forecasts.

In parallel, many studies have argued against the existence of such superhorizon loop corrections~\cite{Riotto:2023hoz,Tada:2023rgp,Fumagalli:2023zzl}.
These counterarguments are rooted in two distinct approaches: intricate one-loop calculations that incorporate backreaction~\cite{Inomata:2024lud,Inomata:2025bqw,Fang:2025vhi,Inomata:2025pqa,Braglia:2025cee,Braglia:2025qrb}, and arguments based on the consequences of spatial dilatation symmetry~\cite{Fumagalli:2024jzz,Kawaguchi:2024rsv}. 
However, these complex calculations fail to provide a unifying principle for the observed cancellations, leaving it unclear if conservation is merely a one-loop coincidence.
 In quantum field theory, such finely-tuned cancellations are hallmarks of a Ward identity. This strongly suggests that a hidden symmetry may be at work. We are thus led to ask:\textbf{ Can a Ward identity derived directly from symmetry impose non-perturbative constraints on the evolution of superhorizon curvature perturbations?}

In this letter, We illustrate that when backreaction is properly incorporated, perturbation theory naturally satisfies a symmetry due to the redundant freedom in background separation.
To establish the implications of this symmetry, we first systematically construct a framework for disentangling the classical background from the full quantum system—one that naturally embeds the counterterms governing the one-point correlation function. By introducing a generalization of one-particle irreducible (1-PI) diagrams for inflationary spacetimes, we analyze the diagrammatic structure of the power spectrum up to all loop orders. This analysis isolates the dominant infrared (IR) contributions, revealing that the IR behavior of the power spectrum is governed by the zero-mode.
This enables us to leverage the full potency of the symmetry, whose associated Ward identity directly constrains the evolution of the zero-mode and facilitates the construction of an all-order conserved quantity corresponding to the curvature perturbation.

\textit{\textbf{Decomposition of background and perturbations.}} 
We consider a single field inflation model with potential $V(\phi)$ in the spatially flat gauge. The potential is constructed to generate an inflationary evolution featuring an intermediate phase between two slow-roll (SR) periods, namely SR\,-intermediate period-\,SR. We assume that the first SR parameter $\epsilon$ remains small throughout the entire inflationary process, which is applicable to both ultra-slow-roll and parametric resonance scenarios~\cite{Inomata:2024lud,Inomata:2022yte,Cai:2019bmk}. With this set-up, the lapse and shift are suppressed by $\epsilon$, thus justifying the decoupling limit. The dynamics are well described by the following action for the scalar field~\cite{Inomata:2024lud,Inomata:2025bqw}:
\begin{align}
S=\int dt\,d^3x\,a^3\left[\Bigl(\frac12\dot\phi^2 - \frac{(\partial_i\phi)^2}{2a^2}\Bigr) - V(\phi)\right]\,.
\end{align}
It is important to emphasize that under these conditions, interaction effects are non-negligible solely during the intermediate stage.

In inflationary field theory, it is generally assumed that the deviation of the field from a spatially homogeneous background is perturbative. Therefore, the common practice is to subtract a classical field background from the full field operator $\delta\hat{\phi}\equiv\hat{\phi}-\bar{\phi}$ and then use perturbation theory to compute the correlation functions of the perturbations. This subtracted classical field possesses considerable arbitrariness; as long as the redefined field remains suitable for perturbation theory, the classical field can be adjusted arbitrarily. Here, we adopt the following definition for simplicity
\begin{align}
\bigl(a^3\dot{\bar\phi}\bigr)^{\vcenter{\hbox{$\cdot$}}}
= -\,a^3\sum_{n=1}^{\infty}\frac{1}{(n-1)!}\,V^{(n)}(\bar\phi)\,
\langle\delta\phi^{\,n-1}\rangle\,,\label{EoMbarphi}
\end{align}
where $V^{(n)}(\bar\phi)=d^{n}V/d\phi^n|_{\bar\phi}$. To calculate the correlation functions of perturbations, we isolate the action for the perturbation $\delta\phi$, denoted as $S_{\delta\phi}$~\cite{Kristiano:2025ajj}:
\begin{equation}
\begin{aligned}
S_{\delta\phi}
&= \int dt\,d^3x\,a^3\Bigl(\frac12\,\delta\dot\phi^2
   - \frac12\,\frac{(\partial_i\delta\phi)^2}{a^2}\Bigr)
   + \bigl(a^3\dot{\bar{\phi}}\,\delta\phi\bigr)^{\vcenter{\hbox{$\cdot$}}} \\
   &\quad - (a^3\dot{\bar\phi})^{\vcenter{\hbox{$\cdot$}}}\,\delta\phi
   - a^3\sum_{n=1}^\infty\frac{1}{n!}\,V^{(n)}(\bar\phi)\,\delta\phi^n\,.
\end{aligned}
\end{equation}
The above action contains two notable terms arising from the integration by parts. The total derivative term, which evaluates on the time boundary, can be discarded as it does not contain the time derivative of $\delta\phi$~\cite{Inomata:2025bqw}. The one-point interaction term, $(a^3\dot{\bar\phi})^{\vcenter{\hbox{$\cdot$}}}\,\delta\phi$, acts as a counter-term that ensures the one-point correlation function $\langle\delta\phi\rangle$ vanishes, i.e., $\langle\delta\phi\rangle=0$ even after the interactions have been opened.

\textit{\textbf{IR structures of the Feynman diagrams.}} 
To compute correlation functions perturbatively, we work within the in-in formalism. The analysis begins with the free field operator in the interaction picture, $\delta\hat{\phi}^{(1)}$.
Crucially, we separate the zero-mode ($k=0$) from the finite-momentum modes ($k \neq 0$):
\begin{equation}
\begin{aligned}
    \delta\hat{\phi}^{(1)}(\boldsymbol{x},t) 
&=u_{0}(t)\hat{a}_{\boldsymbol{0} }+ u_{0}^{*}(t)\hat{a}^{\dagger}_{\boldsymbol{0}}+\\
&\int_{k\neq 0}\frac{\mathrm{d}^3k}{(2\pi)^3} \mathrm{e}^{i\boldsymbol{k}\cdot\boldsymbol{x}} \left[ u_{k}(t)\hat{a}_{\boldsymbol{k}} + u_{k}^{*}(t)\hat{a}^{\dagger}_{-\boldsymbol{k}}\right]\,,
\end{aligned}
\end{equation}
This separation is essential as their respective mode functions, $u_0(t)$ and $u_k(t)$, obey different equations of motion (Eom) due to the absence of a spatial gradient term for the zero-mode
\begin{align}
\bigl(a^3\dot{u}_0\bigr)^{\vcenter{\hbox{$\cdot$}}} + a^3V^{(2)}(\bar\phi)u_0 &= 0\,, \\
\bigl(a^3\dot{u}_k\bigr)^{\vcenter{\hbox{$\cdot$}}} + a^3V^{(2)}(\bar\phi)u_k + a k^2 u_k &= 0 \quad (k\neq 0)\,.
\end{align}
We consider sufficiently small k-modes. During the early SR phase, the Eom yields a frozen solution proportional to that of the zero-mode, allowing us to define their ratio as $C\equiv u_k(t_i)/u_0(t_i)$. After entering the USR stage, modes in the infrared limit $(k\rightarrow0)$ still satisfy the same equation as the zero-mode, so the ratio $C$ determined by the initial conditions remains a constant throughout the evolution.


Now we step forward to the non-linear corrections to the dimensionless power spectrum $\mathcal{P}(k)$, which is defined as $
    (2\pi)^3\delta^{(3)}(\boldsymbol{k}+\boldsymbol{p})\mathcal{P}(k) \equiv {k^3}/{2\pi^2}\langle\delta \phi_{\boldsymbol{p}}\delta \phi_{\boldsymbol{k}}\rangle\,.$
Here we assumed that the perturbative conditions are not violated, so we can compute correlation functions using Feynman diagrams and then analyze their structure. According to momentum conservation, the power spectrum of finite $k$ includes only connected Feynman diagrams. In contrast, the zero-mode correlator $\langle\delta\phi_0\delta\phi_0\rangle$ contains connected and disconnected parts, $\langle\delta\phi_0\delta\phi_0\rangle = \langle\delta\phi_0\delta\phi_0\rangle_c + \langle\delta\phi_0\rangle \langle\delta\phi_0\rangle,$
where the subscript `$c$' denotes the connected part of the correlator. 
We found that the disconnected parts can always be factorized into forms like $\langle\delta\phi_0\rangle \langle\delta\phi_0\rangle$
As established previously, the counter-terms in the action ensure that the one-point function vanishes, i.e., $ \langle\delta\phi_0\rangle = 0$. Consequently, the disconnected part of the two-point function is zero.

This leads to a key conclusion: the full zero-mode correlator is equal to its connected part, $\langle\delta\phi_0\delta\phi_0\rangle = \langle\delta\phi_0\delta\phi_0\rangle_c$. Therefore, both the quantity relevant for the power spectrum, $\langle\delta\phi_{-\boldsymbol{k}}\delta\phi_{\boldsymbol{k}}\rangle_c$, and the full zero-mode correlator, $\langle\delta\phi_0\delta\phi_0\rangle$, are described by the same set of diagrams: the sum of all connected diagrams with two external points.

To analyze the IR structure of the two-point function $\ev{\delta \phi_{-\boldsymbol{k}}\delta \phi_{\boldsymbol{k}}}$, we must first recognize that its diagrammatic expansion consists of two components, distinguished by their origin and IR behavior.

The first component, the propagator, arises from the Dyson series expansion of the Heisenberg picture operator, which expresses $\delta\phi_H$ as a sum of nested commutators with the interaction Hamiltonian $H_I$ 
\cite{Wang:2013zva}:
\begin{equation}
\begin{aligned}
    \delta\hat{\phi}_H=&\sum_{n=0}^\infty i^n
    \int_{t_i}^{t}\frac{dt_1}{a(t_1)}\cdots\int_{t_i}^{t_{n-1}}\frac{dt_n}{a(t_n)}\,\\
    &\times[H_I(t_n),\dots,[H_I(t_1),\delta\hat{\phi}_I]\dots] ~.
\end{aligned}
\end{equation}
These nested commutators systematically yields retarded Green's functions, $G_k(t, t')$, i.e.,
\begin{equation}
    [\delta\hat{\phi}_{\boldsymbol{p}}(t'),\delta\hat{\phi}_{\boldsymbol{q}}(t'')]=W(t')\,G_q(t'',t')\,(2\pi)^3\delta^3(\boldsymbol{p}+\boldsymbol{q})\,,
\end{equation}
where $W(t)$ denotes the Wronskian.
In Feynman diagrams, these Green's functions are represented as arrowed lines (a-lines), signifying causality. Their crucial property is that they are regular in the IR limit: $\lim_{k\to0}G_k(t,t') = G_0(t,t')$ \cite{Inomata:2025bqw}.

The second component, the correlator, emerges when taking expectation values, as this is equivalent to performing all possible Wick contractions of free field operators. 
According to Wick's theorem, these fields are contracted in pairs, yielding products of mode functions like $u_p(t') u_p^*(t'')$, which are exactly free correlators.
We define these wick contractions ($u_p u_p^*$) as non-arrowed lines (na-lines) in diagrams. In stark contrast to a-lines, these na-lines are singular in the IR limit: $u_p(t') u_p^*(t'') \sim {\mathcal{P}_{1}(p)}/{p^3} \quad \text{as} \quad p \to 0$, where $\mathcal{P}_{1}$ denotes tree level spectrum.

To analyze how these two distinct components contribute to the dimensionless power spectrum, $\mathcal{P}(k)$, we first classify all connected diagrams based on their topology. Any general connected diagram can always be categorized into two types:
\begin{itemize}
    \item \textbf{Reducible (or Cuttable) Diagrams}: Those that can be disconnected into two separate parts by cutting a single \textbf{na-line}. 
    \item \textbf{Irreducible (or Non-Cuttable) Diagrams}: Those that remain connected after cutting any single na-line. 
\end{itemize}
This classification is a natural generalization of the standard concept of one-particle irreducible (1-PI) diagrams. As we will now demonstrate, this classification precisely separates the diagrams by their IR behavior, with only the reducible diagrams contributing in the IR limit.

First, for an irreducible diagram (non-cuttable), any na-line must by definition be part of a closed loop. Since we assume all loop integrals are regularized and remains analytic functions of the external momentum $k$, their values approaches a finite constant as $k\to 0$. Any external dependence on $k$ can only come from a-lines, which are themselves regular. The total value of an irreducible diagram is therefore a regular function of $k$.
When these regular behaviors are inserted into the definition of the power spectrum, its contribution is suppressed by the $k^3$ prefactor and vanishes in the deep IR limit:
\begin{align}
    \mathcal{P}(k) \propto k^3 \times (\text{finite constant}) \xrightarrow{k \to 0} 0\,.
\end{align}

Next, for a reducible (cuttable) diagram, there exists a single na-line that acts as a bridge connecting the two sub-diagrams. By momentum conservation, it is guaranteed that this specific na-line must carry the full external momentum, $k$, which endows the diagram with a $\mathcal{P}_{1}(k)k^{-3}$ singularity in the IR limit. It is worth noting a key structural rule from the in-in formalism: na-lines only connect fields originating from different Heisenberg operators. This has a direct topological consequence: any reducible diagram must contain precisely \textit{one} such cuttable na-line. Due to this singular component, the diagram's contribution to the correlator will scale as $\mathcal{P}_{1}(k)k^{-3}$ for $k\to 0$. In this case, the $k^3$ prefactor in the power spectrum definition can be canceled, leading to a contribution proportional to the tree level diagram:
\begin{align}
    \mathcal{P}(k) \propto k^3 (C \mathcal{P}_{1}(k)k^{-3} + O(1)) \xrightarrow{k \to 0} C\mathcal{P}_{1}(k)\,.
\end{align}
This analysis leads to a powerful and predictive conclusion: only reducible diagrams can provide a non-vanishing contribution to the power spectrum in the $k \to 0$ limit. This classification will be the cornerstone of our all-loop analysis.


 As a concrete illustration, We show two diagrams with different structures. A diagram with the topology of~\ref{fig:two_loops}(a) is irreducible and thus IR-suppressed, while a diagram like~\ref{fig:two_loops}(b) is reducible and provides a leading-order contribution.

\begin{figure}[ht]
  \centering
  \begin{minipage}[c]{0.4\linewidth}
    \centering
    \begin{tikzpicture}[line width=1pt,baseline=(v.base),scale=0.7]
      \coordinate (i) at (-1.4,0);
      \coordinate (v) at (0,0);
      \coordinate (o) at (1.4,0);
      \coordinate (a) at (-1.2,0);
      \coordinate (b) at (1.2,0);
      \def\r{0.8}
      \coordinate (c) at (0,-1.2);
      \draw (i) -- (o);
      \draw (c) circle (\r);
      \coordinate (f) at ({-\r/sqrt(2)}, {-1.2+\r/sqrt(2)});
      \coordinate (g) at ({ \r/sqrt(2)}, {-1.2+\r/sqrt(2)});
      \coordinate (h) at ({-\r/sqrt(2)}, {-1.2-\r/sqrt(2)});
      \coordinate (j) at ({ \r/sqrt(2)}, {-1.2-\r/sqrt(2)});
      \draw[thick] (a) -- (f) node[midway, sloped, allow upside down]{$<$};
      \draw[thick] (b) -- (g) node[midway, sloped, allow upside down]{$<$};
      \draw[thick] (c) ++(-\r,0) arc (170:180:\r) node[midway,sloped,allow upside down]{$<$};
      \draw[thick] (c) ++( \r,0) arc (170:180:\r) node[midway,sloped,allow upside down]{$<$};
      \draw[thick] (h) ..controls (-0.2,-1.2) and (0.2,-1.2).. (j);
    \end{tikzpicture}
    \caption*{(a) Irreducible 2-loop}
    \label{fig:irr_two_loop}
  \end{minipage}%
  \hspace{0.04\linewidth}
  \begin{minipage}[c]{0.4\linewidth}
    \centering
    \begin{tikzpicture}[line width=1pt,baseline=(v.base),scale=0.7]
      \coordinate (i) at (-1.4,0);
      \coordinate (v) at (0,0);
      \coordinate (o) at (1.4,0);
      \coordinate (a) at (-1.2,0);
      \coordinate (b) at (1.2,0);
      \def\r{0.4}
      \coordinate (c) at (-0.6,-1);
      \coordinate (d) at ( 0.6,-1);
      \draw (i) -- (o);
      \draw (c) circle (\r);
      \draw (d) circle (\r);
      \coordinate (f) at ({-0.6-\r/sqrt(2)}, {-1+\r/sqrt(2)});
      \coordinate (g) at ({ 0.6+\r/sqrt(2)}, {-1+\r/sqrt(2)});
      \draw[thick] (f) -- (a) node[midway, sloped, allow upside down]{$>$};
      \draw[thick] (g) -- (b) node[midway, sloped, allow upside down]{$>$};
      \draw[thick] (c) ++(0,\r) arc (90:270:\r) node[midway,sloped,allow upside down]{$<$};
      \draw[thick] (d) ++(0,\r) arc (90:-90:\r) node[midway,sloped,allow upside down]{$<$};
      \draw[thick] (c) ++(0,-\r) ..controls (0,-1.7).. (0.6,-1-\r);
    \end{tikzpicture}
    \caption*{(b) Reducible 2-loop}
    \label{fig:red_two_loop}
  \end{minipage}
  \caption{Examples of two-loop diagrams: (a) irreducible and (b) reducible.}
  \label{fig:two_loops}
\end{figure}
The structure of reducible diagrams allows us to relate the finite-momentum power spectrum to the zero-mode correlator. Given the established relationship between the mode functions, $u_k(t') u_k^*(t'') = C^2 u_0(t') u_0^*(t'')\,(k\rightarrow 0)$, the leading IR behavior of any reducible diagram for $\mathcal{P}(k)$ becomes directly proportional to its zero-mode counterpart.

Furthermore, since the single cuttable na-line in the reducible diagrams is the unique bridge connecting fields originating from two different Heisenberg operators, we can thus define a new effective mode function $U_k(t)$, through the relation $\langle\delta\hat{\phi}_{\boldsymbol{k}}(t) \hat{a}^{\dagger}_{\boldsymbol{-k}}\rangle= U_k(t)\left[\hat{a}_{\boldsymbol{k}},\hat{a}^{\dagger}_{\boldsymbol{k}}\right]$\footnote{This definition is both valid for finite momentum and zero modes.}. The IR behavior is governed by the reducible diagrams, and thus is governed by the evolution of $U_k(t)$.

Therefore, the problem of finding the IR limit of the power spectrum nonlinearly is equivalent to determining the evolution of the zero-mode function, $U_0(t)$. This evolution can be constrained non-perturbatively by a powerful symmetry analysis.

\textit{\textbf{Constraints from symmetry.}}
We now introduce a symmetry of the system to non-perturbatively constrain the evolution of this zero-mode function. Consider the following set of transformations, parameterized by an infinitesimal constant $\lambda$
\begin{align}
\tilde{\boldsymbol{x}} = (1 - \lambda)\boldsymbol{x}, \,
\tilde{t} = t + \frac{\lambda}{H}, \, 
\delta\tilde{\phi}(\tilde{\boldsymbol{x}}, \tilde{t}) = \delta\phi(\boldsymbol{x}, t) - \lambda \frac{\dot{\bar{\phi}}}{H}\,.
\end{align}
We examine the transformation of the action under this set of variable substitutions
\begin{align}
\begin{aligned}
S =& \int (1 + \epsilon \lambda)\, d\tilde{t}\, d^3 \tilde{x}\, \tilde{a}^3 \biggl[\frac{1}{2}\biggl[-  \frac{(\tilde{\partial}_i \delta \tilde{\phi})^2}{\tilde{a}^2}\\
& +\biggl( \dot{\bar{\phi}}(\tilde{t}) - \lambda \left(\frac{\dot{\bar{\phi}}}{H} \right)^{\cdot}+ \delta \dot{\tilde{\phi}}  + \lambda \left(\frac{\dot{\bar{\phi}}}{H} \right)^{\cdot}-\epsilon \lambda \delta \dot{\phi} \biggr)^2 
\biggr]  \\
 & - V\biggl( \bar{\phi}(\tilde{t}) - \lambda\frac{\dot{\bar{\phi}}}{H} + \delta \tilde{\phi} + \lambda\frac{\dot{\bar{\phi}}}{H}\biggr)\biggr]\,.
\end{aligned}
\end{align}
in the $\epsilon \to 0$ limit, we immediately notice that the form of the action remains unchanged, which means $S[\delta\phi] \equiv S[\delta \tilde{\phi}]$. 
The Ward identity associated with this symmetry~\cite{Assassi:2012zq,Assassi:2012et} is given by,
\begin{align}
i [\hat{Q}, \delta \hat{\phi}] = - \delta\delta \hat{\phi} ~,\label{wardidentity}
\end{align}
where $\hat{Q}$ is the conserved charge associated with the symmetry, and the field variation $\delta\delta \hat{\phi}$ is given by
\begin{align}
\delta\delta \hat{\phi} = {x^i \partial_i \hat{\phi}} - \frac{
\delta\dot{\hat{\phi}}}{H}-\frac{\dot{\bar{\phi}}}{H}\,.
\end{align}

Consider an eigenstate of field configurations that reads $\delta \hat{\phi}(x) |\delta \phi \rangle = \delta \phi(x) |\delta \phi \rangle$. Taking expectation values of both sides of Eq.~\ref{wardidentity}
and noticing that $\langle \Omega | \delta \delta\hat{\phi} | \Omega \rangle = -{\dot{\bar{\phi}}}/{H}$, the Ward identity thus gives
\begin{equation}\label{Ward}
    \frac{\dot{\bar{\phi}}}{H}(t)=i \int D\delta\phi_{i} \;\left[ \langle \Omega | \hat{Q} | \delta\phi_{i} \rangle \langle \delta\phi_{i} | \delta\hat{\phi} | \Omega \rangle - c.c.\right],
\end{equation}
where we have inserted a complete set of field eigenstates $\ket{\delta\phi_i}$ at an early time $t_i$.

To evaluate the matrix element $\langle \Omega | \hat{Q} | \delta\phi_{i} \rangle$, we will analyze how the vacuum wave functional transforms under the symmetry operation. Our strategy is to compute the wave functional of the transformed eigenstate, $\ket{\Psi}\equiv(1 - i \lambda \hat{Q})\ket{\delta\phi}$, and compare it to the original wave functional, $\langle \Omega | \delta \phi \rangle$.

The vacuum wave functional at early times, when interactions are negligible, is the standard Bunch-Davies Gaussian state~\cite{Hui:2018cag,Maldacena:2002vr,Chen:2017ryl,Hinterbichler:2013dpa}
\begin{align}\label{wave function}
\begin{aligned}
    \langle \Omega | \delta \phi \rangle 
\propto&\,\exp{\left(
 -\frac{1}{2} \epsilon_0(t ) \delta{\phi}_{\bf{0}} \delta{\phi}_{\bf{0}}\right)}\\
 &\exp \left[ \int_{k\neq0} \frac{d^3k}{(2\pi)^3} \left(
 -\frac{1}{2} \epsilon_k(t ) \delta{\phi}_{\bf{k}} \delta{\phi}_{-\bf{k}}
\right) \right]\,.
\end{aligned}
\end{align}

A key step is to determine the properties of the transformed state $\ket{\Psi}$. Using the infinitesimal form of $\tilde{\delta\phi}$ and~\eqref{wardidentity}, one can show that $\ket{\Psi}$ satisfies
\begin{equation}
    \left[ \delta\hat{\phi} \left( (1+\lambda)\boldsymbol{x}, t - \frac{\lambda}{H} \right)-\lambda\frac{\dot{\bar{\phi}}}{H}  \right] \ket{\Psi}
= {\delta\phi}(\boldsymbol{x}) \ket{\Psi}\,,
\end{equation}
which implies that the transformed state $\ket{\Psi}$ is also a field eigenstate, but with a modified eigenvalue $\psi(\boldsymbol{x})$
\begin{equation}
    \delta\hat{\phi}\left( \boldsymbol{x}, t - \frac{\lambda}{H} \right) \ket{\Psi}
= \psi(\boldsymbol{x}) \ket{\Psi}\,,
\end{equation}
where $\psi = \lambda{\dot{\bar{\phi}}}/{H} + \delta {\phi}((1-\lambda)\boldsymbol{x})$.

Evaluating the wave functional requires the Fourier modes of this new eigenvalue, $\psi_{\boldsymbol{k}}$, which can be found to be
\begin{align}
\psi_{\boldsymbol{k}} = (1+3\lambda)\delta\phi_{\boldsymbol{k}(1+\lambda)}\ ,
\quad \psi_0 = \delta\phi_0 + \lambda\frac{\dot{\bar{\phi}}}{H} ~.
\end{align}
We also need to derive the transformation rules of the gaussian kernel. Since these kernels only depend on mode functions, we can conclude from the transformation property of the early time mode functions $u_{k}(t_i)$ that
\begin{align}
\begin{aligned}
    &\epsilon_k\left(t-\frac{\lambda}{H}\right)=(1-3\lambda) \epsilon_{k(1+\lambda)}(t) ~,\\
    &\epsilon_0\left(t-\frac{\lambda}{H}\right) = \epsilon_0(t)\,.
\end{aligned}
\end{align}

With these ingredients, we can assemble the transformed wave functional. By substituting the transformed eigenvalues and kernels into the Gaussian form~\eqref{wave function} and expanding to first order in $\lambda$, we find a simple relation
\begin{align}
\begin{aligned}
&\langle\Omega|\Psi\rangle \propto \exp\left[ -\frac{1}{2}\epsilon_0 \left(t-\frac{\lambda}{H}\right)\psi_0^2  \right]\\
&\exp\left[ \int_{k\neq0} \frac{d^3k}{(2\pi)^3} \left(-\frac{1}{2} \epsilon_k\left(t-\frac{\lambda}{H}\right) \psi_{\mathbf{k}} \psi_{-\mathbf{k}}\right) \right]\\
&=\left(1- \lambda\epsilon_0 \frac{\dot{\bar{\phi}}}{H} \delta\phi_0\right)\langle \Omega | \delta \phi \rangle\,,
\end{aligned}
\end{align}  
which can then be reduced by $\langle\Omega|\delta\phi\rangle$ to yield the expected results
\begin{align}
i\lambda \langle\Omega| Q  |\delta\phi_i\rangle = \lambda \epsilon_0 \frac{\dot{\bar{\phi}}}{H} \delta\phi_0 \langle\Omega|\delta\phi_i\rangle\,.
\end{align}

Substituting our result for the matrix element into~\eqref{Ward}, the Ward identity becomes
\begin{equation}
\begin{aligned}
\frac{\dot{\bar{\phi}}}{H}(t) &= \int D\delta\phi_i \left[\epsilon_0 \frac{\dot{\bar{\phi}}}{H} \delta\phi_0 \langle\Omega|{\delta\phi}_i\rangle \langle\delta\phi_i|\delta\hat{\phi}|\Omega\rangle + c.c.\right] \\
&= \epsilon_0 \frac{\dot{\bar{\phi}}}{H}\langle\Omega|\delta\hat{\phi}_0(t_i)\delta\hat{\phi}_0|\Omega\rangle + c.c\,.
\end{aligned}
\end{equation}
where we have utilized the property $\delta\hat{\phi}_0(t_i)=\int D\delta\phi_i \delta\phi_0 |{\delta\phi}_i\rangle \langle\delta\phi_i|$. Since the Gaussian kernels are directly related to the two-point correlators of $\delta\phi$, i.e.,$\epsilon_0=1/|u_0|^2$ (see appendix \ref{appendix:phase} for details), the above expression simplifies to
\begin{align}
    \left\langle\delta\hat{\phi}_0(t_i)\frac{H \delta\hat{ \phi}_{0}(t)}{\dot{\bar{\phi}}(t)}\right\rangle + c.c. = 2 \left\langle\delta\hat{\phi}_0(t_i)\frac{H \delta\hat{\phi}_0(t_{i})}{\dot{\bar{\phi_i}}(t_i)}\right\rangle\,.
\end{align}
Because $\delta\hat{\phi}_0(t_{i})$ only contains $\hat{a}_{\boldsymbol{0} }$, this equation implies that the Bogoliubov coefficients of $\delta\phi_0(t)$ evolve proportionally to $\dot{\bar{\phi}}/{H}$. Moreover, given that for IR momentum, $U_k(t) 
= C U_0(t),$
 and considering our earlier analysis of the loop diagram structure, we arrive at the final result:
\begin{align}\label{conversed}
\lim_{k\to 0} \frac{H^2 \mathcal{P}(k)}{\dot{\bar{\phi}}^2} \quad \text{is constant}\,.
\end{align}
This quantity corresponds directly to the curvature power spectrum in the sense of the nonlinear $\delta N$ formalism \cite{Inomata:2024lud,Caravano:2025diq,Pi:2022ysn}.
 We arrived at the main result of this work in Eq.~\eqref{conversed}. It shows that the Ward identity can indeed impose strong constraints on the evolution of the two-point correlation function beyond one-loop order directly, without carrying out explicit loop calculations.
 
\textit{\textbf{Conclusions and discussion.}} 
The core conclusion of this paper is that, even in the presence of non‑slow‑roll phases, the evolution of the curvature perturbation spectrum in the super‑horizon limit still satisfies a conservation law originating from symmetry. Provided that we consider the symmetry transformation with background‑fluctuation splitting and assume that this symmetry remains valid for the bare Lagrangian including all renormalization counterterms, the conservation of the curvature perturbation spectrum follows directly from the Ward identity.

Our proof, relates to several interesting new questions for future investigation. First, renormalizing loop divergences inevitably introduces counterterms whose finite parts are typically undetermined, leading to theoretical ambiguities. However, analogous to how the Ward-Takahashi identity in quantum electrodynamics uniquely fixes the photon mass counterterm to preserve gauge invariance, the background-perturbation splitting symmetry tightly constrains the renormalization procedure. The associated Ward identity strictly dictates that the finite coefficients of these counterterms must vanish, thereby prohibiting any non-vanishing loop corrections to the IR power spectrum. We must emphasize, however, that this classical symmetry could potentially suffer from a quantum anomaly. Which means in generic UV completions, the symmetry may be inherently broken at the quantum level. Yet, rather than weakening our framework, this elevates the superhorizon conservation law to a fundamental diagnostic criterion. A large superhorizon loop corrections would unambiguously signal the anomalous breaking of this underlying symmetry, providing a stringent physical probe for new physics beyond the canonical single-field paradigm. Second, a more complete understanding of nonlinear gauge transformation is required. In particular, many papers have derived the action in the decoupling limit using the effective field theory approach within the unitary gauge. Clarifying how the method employed in this letter corresponds to the effective field theory approach is also an important issue.

It is also instructive to connect the symmetry identified in this work with the spatial dilatation symmetry, $\zeta \rightarrow \zeta+\lambda$ in the comoving gauge, where $\lambda$ is an arbitrary constant~\cite{Assassi:2012et,Assassi:2012zq,Hinterbichler:2013dpa,Hui:2018cag,Tanaka:2015aza,Hinterbichler:2012nm,Fumagalli:2024jzz,Kawaguchi:2024rsv}. We argue that both are manifestations of the same underlying principle: the inherent ambiguity in separating a quantum field into its classical background and perturbation. This freedom allows one to shift a homogeneous mode between these two components. While the standard dilatation involves shifting by a constant $\lambda$, the symmetry we uncovered corresponds to a dynamically determined shift proportional to $\dot{\bar{\phi}}/H$. Crucially, this dynamic shift is precisely the growing-mode solution to the zero-mode's equation of motion. This leads to a powerful reinterpretation: the much-debated infrared behavior of inflationary perturbations is not an independent physical effect, but a direct reflection of this theoretical redundancy in the background-perturbation split.

In future work, we will discuss the gauge transformation problem in more detail and elaborate further on the connection between loop‑diagram calculations and symmetry analysis. We will also apply our method to tensor modes in our on-going work. 

\textit{\textbf{Acknowledgements.}} 
We thank Jason Kristiano, Keisuke Inomata, Zhong-Zhi Xianyu, Shi Pi, Misao Sasaki, Takahiro Tanaka, Yuko Urakawa, Sébastien Renaux-Petel and Junichi Yokoyama for useful discussions. This work is supported in part by the National Key Research and Development Program of China (No. 2020YFC2201501), in part by the National Natural Science Foundation of China (No. 12475067 and No. 12235019). CC is supported by National Natural Science Foundation of China (No. 12433002) and Start-up Funds for Doctoral Talents of Jiangsu University of Science and Technology. CC thanks the supports from The Asia Pacific Center for Theoretical Physics, The Center for Theoretical Physics of the Universe at Institute for Basic Science during his visits.

\bibliographystyle{apsrev4-1}
\bibliography{main}
\appendix
\begin{widetext}
\section{The evolution of the zero-mode}
\label{appendix:phase}

To be rigorous, we still need to further discuss the properties of the effective zero mode function$U_0(t)$, particularly the evolution of its phase. We start from the linear order mode function $u_0$, which fulfills 
\begin{align}
    \bigl(a^3\dot{u}_0\bigr)^{\!\Large\cdot}
+ a^3\,V^{(2)}(\bar\phi)\,u_0 = 0. 
\end{align}
Under the SR condition, we have $V^{(2)} \ll H^2$ during the first SR period. Thus the solution of $u_0$ can be approximately as $u_0\sim c+iba^{-3}$. Considering the initial conditions that minimize the energy, both $b$ and $c$ can be chosen as real constants. Assuming the first SR period is sufficiently long, the real part of $u_0$ dominates. Therefore, $u_0 $ effectively becomes a constant real number before the end of the first SR era. Entering the USR epoch afterward, although $u_0$ may not remain frozen, its evolution is dominated by its real part and remain proportional to $\dot{\bar{\phi}}_0/{H}$, where $\bar{\phi}_0$ is the linear order solution of $\bar{\phi}$. 

From the Schrödinger picture perspective, we can analyze the evolution of quantum states. The mode function evolution above corresponds to a squeezed state compressed along the configuration direction. The ``early" wave function mentioned in the main text refers specifically to such a squeezed state during the first SR period. Consequently, the Gaussian kernel of the wave function can be fixed as $\epsilon_0=1/|u_0|^2$ which is a real number. 

We aim to further investigate the phase evolution of $U_0$ after turning on the interaction. Starting at one-loop order, the corrections of $U_0$ are given by
\begin{align}
    \begin{aligned}
&\int^t\mathrm{d} t_1 a V^{(3)} G_0\left(t ; t_1\right)\int \frac{\mathrm{d}^3 k}{(2 \pi)^3} \int^{t_1} \mathrm{d} t_2 G_k\left(t_1; t_2\right) a V^{(3)}\operatorname{Re}\left[u_k(t_2) u_k^*\left(t_1\right)\right] u_0(t_2)\\
&+\int^t\mathrm{d} t_1 G_0\left(t ; t_1\right)\frac{a}{2}V^{(4)}\int\frac{\mathrm{d}^3 k}{(2 \pi)^3}|u_k(t_1)|^2u_0(t_1)
    \end{aligned}
\end{align}
where we notice that each part inside the integral is a real number. Thus, the one-loop correction does not alter the phase of $U_0$, which remains real. It is natural to ask whether this property remains correct in higher order corrections. In fact, this can be analyzed through node structure and commutator symmetry. There are three possible origins of the imaginary parts
\begin{itemize}
    \item The imaginary unit $i$ in the interaction picture evolution operator
    \item The purely imaginary commutators of field operators
    \item  The Wick contractions (na-lines) of operators at different times which are complex numbers
\end{itemize}
The imaginary contributions from the first two sources cancel out because in the commutator-form expression, the n-th order term contains n commutators and is multiplied by the coefficient $i^n$. Their product yields a purely real result.

Proving that the Wick contraction parts are also real is somewhat non-trivial. We begin with the one-loop order term, which contains a commutator of the following structure
\begin{align}
\left[\delta\phi_2^3,\left[\delta\phi_1^3,\delta\phi_q\right]\right]=3\left[\delta\phi_2^3,\delta\phi_1^2\right]\left[\delta\phi_1,\delta\phi_q\right]=3\left[\delta\phi_1,\delta\phi_q\right]\sum_{m=0}^{2}\delta\phi_2^m\left[\delta\phi_2,\delta\phi_1^2\right]\delta\phi_2^{2-m}
\end{align}
From the symmetry of this commutator, we find that for every possible Wick contraction, there exists a conjugate term which meets the requirement that all contractions within $\delta\hat{\phi}_{\boldsymbol{q}}(t)$ are in the opposite direction to those in the original term. For instance, $u_p(t') u_p^*(t'')$ corresponds to $u_p(t'') u_p^*(t')$ in conjugate terms, causing their imaginary parts to cancel and leaving only $\operatorname{Re}u_p(t'') u_p^*(t')$ in the final expressions. The same reasoning applies to higher-order contributions. The structure of the commutators now becomes
\begin{align}
\left[\delta\phi_i^n,\left[\delta\phi_j^m,\dots\right]\right]=\sum_{p=0}^{n}\delta\phi_i^p\left[\delta\phi_i,\left[\delta\phi_j^m,\dots\right]\right]\delta\phi_i^{n-1-p},
\end{align}
thus for every possible Wick contraction that contains $\delta\phi_i$, we can find its conjugate term. This process proceeds layer by layer, leaving  only the real parts of these contractions. Thus, we conclude that for the mode function $U_0$, turning on the interaction only changes its modulus but not its phase. This conclusion plays a crucial role in our proof.
$$\langle \delta\phi^2(x,\tau)\rangle=\int_0^\infty \mathbf{d}\ln{k} \,P_k=\int_0^\infty \mathbf{d}\ln{\tilde{k}} \,P_{\tilde{k}(1+\lambda)}=\langle \delta\phi^2(x,(1+\lambda)\tau)\rangle
$$
\end{widetext}
\end{document}